\documentclass[tightenlines,a4paper,twocolumn,floatfix,aps,showpacs,prd,epsf]{revtex4}
\usepackage{graphicx}
\usepackage{epsfig}
\usepackage{dcolumn}
\usepackage{bm}
\def\dis{\displaystyle}
\def\beq{\begin{equation}}
\def\eeq{\end{equation}}
\def\barr{\begin{array}}
\def\earr{\end{array}}
\def\shat{\hat s}
\def\that{\hat t}
\def\uhat{\hat u}
\def\gsim{\buildrel{\scriptscriptstyle >}\over{\scriptscriptstyle\sim}}

\begin{document}

\title{Search for Excited Quarks in 
$q\bar{q} \rightarrow \gamma\gamma$  at the LHC}
\author{Satyaki Bhattacharya\footnote{Email Address: bhattacharya.satyaki@gmail.com}, Sushil S. Chauhan \footnote{Email Address: sushil@fnal.gov}, Brajesh C. Choudhary\footnote{Email Address: brajesh@fnal.gov } and Debajyoti Choudhury\footnote{Email Address: debchou@physics.du.ac.in}}
\affiliation{ \indent Department of Physics and Astrophysics, University of Delhi, Delhi 11007, India.}

\begin{abstract}
If quarks are composite particles, then excited states are
expected to play a r\^ole in the Large Hadron Collider phenomena.
Concentrating on virtual effects, and using a 
large part of the CMS
detection criteria, 
we present here a realistic examination of their 
effect in diphoton production at the LHC. 
For various luminosities, we present the $99 \% $
confidence limit (CL) achievable in $\Lambda-M_{q*}$ parameter space
where $\Lambda$ is the compositeness scale and $M_{q^*}$ the mass
of the state. For a $q^*$ of mass 0.5 TeV, $\Lambda \leq 1.55 \, (2.95)$
can be excluded at $99\%$ CL with 30 (200)${\rm fb}^{-1}$ 
integrated luminosity. 
\end{abstract}
\pacs{12.60.Rc, 13.40.-f, 13.85.Qk}
\maketitle
\section{Introduction}
    \label{sec:introd}
The replication of fermion families alongwith the mass hierarchies and
mixings has led one to speculate 
about the possibility of quark-lepton compositeness, namely that 
the Standard Model (SM) fermions are not elementary at all.  The
fundamental matter constituents in such theories, very often termed
{\em preon}s\cite{preon}, experience an hitherto unknown force on
account of an asymptotically free but confining gauge
interaction\cite{preon_binding}, which would become very strong at a
characteristic scale $\Lambda$, thereby leading to bound states
(composites) which are to be identified as quarks and leptons.  In
most such models\cite{comp_mod, Additional_comp}, quarks and leptons
share at least some common constituents.

If this hypothesis were to be true, it is possible, indeed probable,
that excited states of fermions exist at a mass scale comparable to
the dynamics of the new binding force.  In the simplest
phenomenological models \cite{hagi_baur_boud}, the excited fermions
are assumed to have both spin and isospin 1/2 and to have both their
left- and right-handed components in weak isodoublets ({\em i.e.}\
they are vector-like).  Since these interact with the SM particles,
they may be produced at high-energy colliders and would decay back,
radiatively, into an ordinary fermion and a gauge boson (photon,
$W$, $Z$ or gluon).  Pair production of charged excited fermions could
proceed via $s$-channel ($\gamma$ and $Z$) exchanges in $e^+ e^-$
collisions, while for excited neutrinos only $Z$ exchange
contributes. Although $t$-channel diagrams are also possible,
they generally give a negligible contribution to the overall pair
production cross-section on account of the smallness of the
cross-couplings~\cite{hagi_baur_boud}.  However, this very same
interaction between the excited state, its SM counterpart and a gauge
boson may be used to singly produce such states (through both
$s$- and $t$-channel diagrams).  The four LEP collaborations have used
these (and other) modes to essentially rule out such excitations
almost upto the kinematically allowed range~\cite{LEPresults}.  At the
{\sc hera}, on the other hand, both excited leptons and quarks may be
produced singly through $t$-channel diagrams and these processes have
been looked at without any positive results~\cite{HERA}.

At the Tevatron, one may either pair-produce the excited quarks
(primarily through gauge couplings) or produce them singly via
quark-gluon fusion, provided the $q^* q g$ coupling strength is
significant. A striking signal of the latter would be an enhancement
in the dijet production rate with a peak in the invariant-mass
distribution.  
Whereas the D$\O$ collaboration has also excluded the mass region 200 GeV
$< M_{q*} <$ 720 GeV for excited quarks decaying to two jets
\cite{d0_m}, the CDF collaboration considered a multitude of decay
channels, thereby excluding the mass range of 80 GeV $< M_{q*} <$ 570 GeV 
\cite{cdf_m1, cdf_m2}.

The presence of such particles would change the phenomenology even if
they were too heavy to be produced.  Since the confining force
mediates interactions between the constituents, it stands to reason
that these, in turn, would lead to interactions between quarks and
leptons that go beyond those existing within the SM. Well below the
scale $\Lambda$, such interactions would likely be manifested through
an effective four fermion contact interaction~\cite{cont_inter,majhi}
term that is an invariant under the SM gauge group.  The D$\O$ and the
CDF experiments at the Tevatron have searched extensively for excited
quarks decaying to different final states as predicted by various
models, with the negative results translating to lower bounds on
compositeness scale $\Lambda$. The D$\O$ collaboration has put a lower
bound of $\Lambda \geq $ 2.0 TeV at 95$\%$ CL from an analysis
of dijet production \cite{d0_dijet}.  
 The CDF collaboration has also put a lower limit of $\Lambda \geq
2.81$ GeV at 95$\%$ CL studying the $q\bar{q}\rightarrow e\nu$ 
process\cite{cdf_W}. From a phenomenological study of flavor independent
contact interaction for the diphoton final state, the lower bound for the
LHC has been estimated to be $\Lambda_{\pm} >$ 2.88 (3.24) TeV at
95$\%$ CL for an integrated luminosity of 100 (200)
$fb^{-1}$~\cite{diphoton}.

As can be readily appreciated, the different production modes (and 
decay channels, wherever applicable) probe different aspects of the 
effective theory that governs the low energy interactions of these 
excited states. In this paper, we seek to concentrate on one such 
property namely the trilinear 
coupling of the excited quark to its SM counterpart 
and the photon. To be more precise, rather than seeking to actually produce 
these excited states, we would like to investigate their r\^ole in 
photon pair production at the LHC. Analogous to the 
process $e^+ e^- \rightarrow \gamma\gamma(\gamma)$ used to
probe compositeness at LEP, such an exercise would complement the excited
quark direct searches for the mass region above the kinematical 
threshold. Since diphoton production is both a very simple 
final state and likely to be 
well-studied at the LHC, it is of interest to see how well can this 
mode probe compositeness.                                                           

The rest of the paper is organized as follows. In the next section, we
discuss the effective Lagrangian for the theory under consideration
and the new physics contribution to diphoton production. In section
III we discuss various SM backgrounds for the signal. In sections IV
and V respectively, we describe the event generation and photon
candidate reconstruction. Isolation study for photon is discussed in
section VI. Confidence limit calculations and results are presented in
sections VII and VIII respectively. The systematics is discussed in
section IX, and in the last section we summarize this analysis with
our conclusions.

\section{Excited quark contribution to diphoton production}

As our interest is not in the production of the excited states, but 
rather on their contribution to the diphoton rates at a hadronic collider, it 
suffices to consider only the relevant parts of the Lagrangian, namely 
the magnetic transition between ordinary and excited states. In general, 
it is often parametrized by
\begin{equation}
{\mathcal L}_{f^*f} = \frac{1}{2 \, \Lambda}\bar{f^*_R} \, \sigma^{\mu\nu}
\left[\sum_i g_i \; c_i \; T_i^a \; G^a_{i \, \mu\nu} \right] f_L + h.c.,
   \label{eq:Lagrangian}
\end{equation}
where the index $i$ runs over the three SM gauge groups, viz. $SU(3)$, 
$SU(2)$ and $U(1)$ and $g_i$, $G^a_{i \, \mu\nu}$ and $T_i^a$ are the 
corresponding gauge couplings, field strength tensors  
and generators respectively. The 
dimensionless constants $c_i$ are, {\em a priori}, unknown 
and presumably  of order unity. Clearly, the phenomenology would 
depend considerably on the ratios of the constants $c_i$. For example, 
electromagnetic couplings (and hence such decays) of such fermions are 
forbidden if $c_2 = e_f \, c_1$. Thus, the search strategies would depend 
crucially on the strengths of these couplings. 

A further point needs to be noted here. In the event of any 
one of the $c_i'$s dominating the others, the cross section for any 
process governed by the Lagrangian above would scale as some 
power of the ratio $c_i / \Lambda$. Thus, in such a case, 
it makes sense to eliminate $c_i$ altogether in favour of the unknown 
scale $\Lambda$. Furthermore, with the Lagrangian of eq.(\ref{eq:Lagrangian})
being a higher dimensional operator, the cross sections would typically 
grow with the center of mass energy, consequently violating unitarity. This 
is not unexpected in an effective theory as the term in  eq.(\ref{eq:Lagrangian}) is only the first term and the loss of unitarity, to a given order, is  
presumably cured once suitable higher dimensional operators are included. 
An equivalent way to achieve the same goal is to consider the $c_i$ to 
be form factors rather than constants. To this end, we shall consider 
the $q^* q \gamma$ vertex to be given by 
\begin{equation}
\overline{q^*} \, q \, \gamma_\mu (p) \quad : 
\qquad \frac{e}{\Lambda} \; \left(1 + \frac{Q^2}{\Lambda^2}\right)^{-n} \;
             \sigma_{\mu \nu} \; p^\nu
   \label{eq:vertex}
\end{equation}
where $Q$ denotes a relevant momentum transfer. It can be 
checked that, for $Q^2 = s$, unitarity is restored as long as the 
constant $n \geq 1$. In the rest of our analysis, we shall confine 
ourselves to a discussion of $n = 1$. While this might seem to be 
an optimistic choice, it is not quite so. 
As can be readily appreciated, such a form factor 
plays a non-negligible r\^ole only when $Q^2 \gsim \Lambda^2$. Since, at 
the LHC, we shall prove to be sensitive to $\Lambda$ of the order of a few 
TeVs, clearly the form factor plays only a marginal r\^ole in the 
determination of the sensitivity reach.

\begin{figure}[htbp]
\scalebox{0.30}{\includegraphics{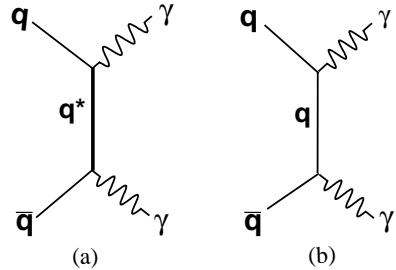}}                                                                               
\caption{\label{fig1}Production process for diphoton final state (a) Excited 
quark mediated (b) SM production.}
\label{fig:fynman}
\end{figure}

With the introduction of the new (flavour-diagonal) 
vertex as in eq.(\ref{eq:vertex}), the 
process $q \bar q \to \gamma \gamma$ acquires a new contribution as 
portrayed in Fig.\ref{fig:fynman}. The differential cross section for 
the partonic process now reads
\begin{equation}
\barr{rcl}
\dis \frac{d \sigma}{d \that} 
& = & \dis
\frac{\pi \, \alpha^2}{3 \, \shat^2} \, 
        \left[
 e_q^4 \, \left( \frac{\uhat}{\that} + \frac{\that}{\uhat} \right)
         - \, \frac{2 \, e_q^2}{\Omega^2} \, 
	 \left( \frac{\that^2}{\hat T} \, 
            + \frac{\uhat^2}{\hat U} \, \right)
   \right.
\\[3ex]
& + & \dis \hspace*{0em}
\left.
      \frac{1}{\Omega^4} \,
   \Bigg\{ 
     \that \, \uhat \, \left(
             \frac{\that^2}{\hat T^2} \, 
            + \frac{\uhat^2}{\hat U^2} \, 
              \right)
+    M_{q^*}^2 \, \shat \, \left(
             \frac{\that}{\hat T} \, 
            + \frac{\uhat}{\hat U} \, 
              \right)^2
\Bigg\} 
\right]
\\[3ex]

\Omega & \equiv & \dis \Lambda \, 
       \left( 1 + \, \frac{\shat}{\Lambda^2} \right)^{n}
\\[3ex]
\hat T & \equiv & \that - M_{q^*}^2
\qquad
\qquad \quad \hat U  \equiv  \uhat - M_{q^*}^2
\earr
\label{eq:matrix}
\end{equation}

where the SM result is recovered in the limit $\Lambda \to \infty$. The 
new physics contribution to the differential cross section thus depends 
on only two parameters, namely
$\Lambda$ and the mass of the excited state $M_{q^*}$. 
For simplicity, we assume these to be flavour-independent (within a 
generation, it obviously has to be so). For  eq.(\ref{eq:Lagrangian})
to make sense as an effective Lagrangian, the masses have to be 
less than $\Lambda$ (Ref.\cite{Hasenfratz:1987tk} requires that 
$M_{q^*} < \Lambda / \sqrt{2}$).

\begin{figure}[htbp]
\scalebox{0.45}{\includegraphics{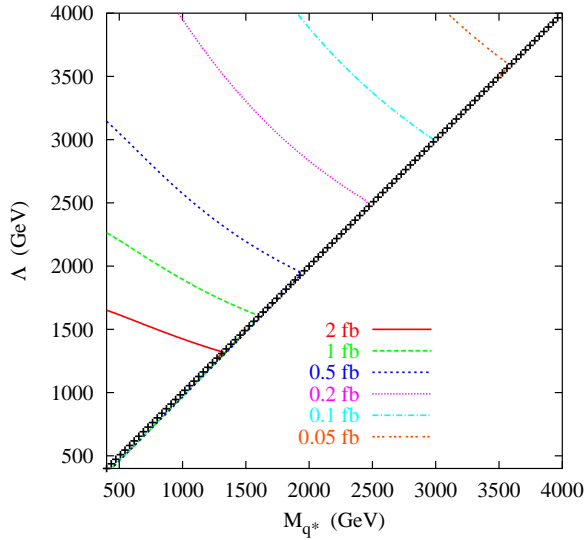}}
\caption{The contribution of new physics to the total cross section.}
\label{fig:new} 
\end{figure}

\indent  In Fig.~\ref{fig:new}, we present the additional contribution to the total diphoton cross section accruing from the new physics terms in eq.(3). Note that, unlike the QED contribution, this additional contribution does not suffer from collinear singularities. 

Contrary to the case of actual production (and subsequent decay) of the 
excited state, the case under consideration is not associated with any 
resonant peak. Nevertheless, the presence of the new contribution 
preferentially enriches the large invariant mass end of the diphoton 
spectrum. The exchange of a large mass particle in the $t$-- and $u$--channels
naturally enhances the high-$p_T$ photon sample. To improve the signal to 
noise ratio, we must then concentrate on such a phase-space 
restricted subset of the final state. 

As can be gleaned from a cursory examination of eq.(3), the  aforementioned dependence of the new contribution on the  photon $p_T$ is not as extreme as that for the QED contribution. Thus, the imposition of such cuts as we will discuss later would not drastically change the shape of the iso-cross section contours as depicted in Fig~\ref{fig:new}. Consequently, the exclusion contours that we shall finally obtain would bear considerable similarity with those in Fig~\ref{fig:new}.

\section{Background}
     \label{sec:bkgd}
Standard Model processes, understandably, 
produce a large background to the diphoton final state. 
The background can be divided into two
categories: 
\begin{itemize}
\item where two prompt photons are produced in the 
      (hard) subprocess itself,  and
\item in a $\gamma +jet$ sample,  a jet with a large
     electromagnetic fraction (e.g, $\pi^{0},\omega, \eta$ etc.) fakes a photon or a hard photon is produced
     in the process of fragmentation.
\end{itemize}
The first  category is 
dominated by the Born-level process 
$q\bar{q} \rightarrow \gamma \gamma$. 
An additional source of the diphoton final state 
is provided by the $gg \rightarrow \gamma\gamma$ process induced
by a box diagram.  
Although the cross-section for this process is
relatively small compared to the Born production (in fact, much 
smaller if very forward photons were to be included)
the much larger $gg$
luminosity at the LHC energies implies that $gg \rightarrow
\gamma\gamma$ can be quite important. Indeed, even after imposing our 
selection criteria (to be discussed later) of moderately low rapidities 
and high transverse momenta for the photons, the $gg$-initiated contribution 
is approximately 6.8\% of the Born contribution (see Table I).
\begin{table}[!h]
 \caption{Various SM
cross-sections for $\hat P_{T}\geq$190 GeV and 
$|\eta|< $2.7 at $\sqrt{s}=$14 TeV. $\hat P_T$, the CKIN(3) 
parameter in PYTHIA, is the $P_{T}$ of the 
outgoing partons in center of momentum frame in a 
$2 \rightarrow 2$ hard scattering process. }  
\begin{ruledtabular} \begin{tabular}{cc} Process  & Cross-Section (fb) \\
\hline
$\gamma+jet$ & 48970\\
 $q\bar{q}\rightarrow\gamma\gamma$ (Born) & 76.05\\
 $gg \rightarrow\gamma\gamma$ (Box)   & 5.18  \\
            
\end{tabular}
\end{ruledtabular}
\end{table}


Apart from the Born and box
processes, single photon production processes $qg \rightarrow \gamma q,
\, q \bar{q} \rightarrow \gamma g $ and $gg \rightarrow \gamma g$ where
a jet fakes a photon can be a major source of background. We have
considered all these processes for the background estimation. Although
the probabilty of a jet faking a photon is $\sim 10^{-3}-10^{-4}$, the
cross section for the first two of these hard processes 
($qg \rightarrow \gamma q
\ , q \bar{q} \rightarrow \gamma g $) are larger by a typical factor of 
${\cal O}(\alpha_s / \alpha)$ apart from a single ratio of gluon to 
quark densities, thereby partly recompensing for this
suppression. The third process, viz. $gg \rightarrow \gamma g$, is once again 
box-mediated and significantly smaller than the other two. 
Similar considerations hold for the background from dijet production 
with both jets being identified as photons. While the dijet
cross section is very large, isolation requirements reduce it drastically. 
Even a simple estimate, without a full simulation, shows it to be 
quite unimportant for the physics under investigation.

\section{Monte Carlo simulation \& Cuts}
     \label{sec:monte} 
To generate the signal as well as the background events, we have used
the {\sc pythia} \cite{pythia} event generator wherein the signal
matrix element of Eq.(\ref{eq:matrix}) had been properly incorporated
inside the {\sc pythia} framework. It was also counterchecked with a
parton-level Monte Carlo generator.  We have used the CTEQ5L parton
distributions~\cite{Lai:1999wy}, with a choice of $Q^2 = \hat s$ for
the factorization scale. While generating
events, the multi parton interaction (MPI), initial state radiation
(ISR) and final state radiation (FSR) switches in {\sc pythia} were
kept ``ON''.  

\begin{figure}[htbp]
\scalebox{0.35}{\includegraphics{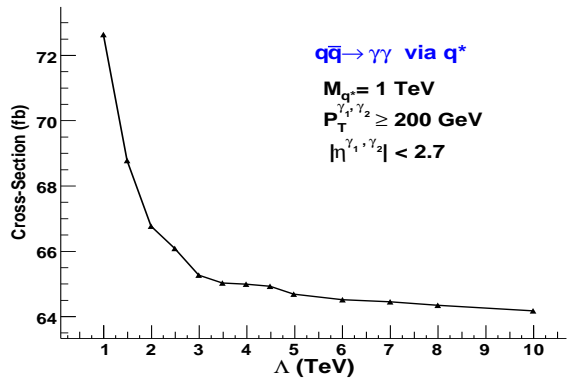}}
\caption{Variation of cross section with $\Lambda$ at $\sqrt{s}=14$
TeV} \label{fig:cs} 
\end{figure}

In view of the fact that the signal events preferentially populate the 
large transverse momentum part of the phase space, events were
generated with $\hat P_{T} \geq$ 190 GeV (CKIN(3) parameter) and 
$|\eta| < 2.7$ respectively. This also rids us of a 
very large fraction of the SM events which, understandably, are peaked 
at small angles to the beam-pipe as also small transverse momenta. 
Fig.~\ref{fig:cs} shows the variation in cross section with
$\Lambda$ for a fixed value of $M_{q*}=$1.0 TeV. Clearly, the variation 
is well-approximated by a $\Lambda^{-2}$ contribution 
superimposed upon a constant (the SM value). This is reflective of the 
fact that, for large $\Lambda$, the new physics contribution is dominated 
by the interference term in Eq.(\ref{eq:matrix}) rather than the pure 
$\Lambda^{-4}$ term . Only if we had imposed harder cuts on the photons, 
would the latter term have dominated (albeit at the cost of reducing event 
numbers and hence the sensitivity). 

It must be noted at this stage that, in the final selection, 
we have used the fiducial volume of the electromagnetic calorimeter of
the CMS detector i.e. $|\eta| < 2.5$ with 1.444 $\leq |\eta| \leq$
1.566 excluded on account of the insensitive region between the barrel and the
endcaps\cite{hybrid}.

\section{Photon Candidate}
    \label{sec:photon}
 Since the SM $\gamma+jet$ and jet-jet production processes form a
significant background to $q\bar{q} \rightarrow \gamma\gamma$ via q*
exchange, it is very important to understand the mechanism of a jet
faking a photon. The identification of a reconstructed object as a
photon candidate depends on the specific design of the detector and
the reconstruction algorithm.  Taking this into consideration, at the
generator level, we have used a clustering algorithm to account for
fake photons arising from jets~\cite{cluster_algo}.
The CMS experiment uses
$ PbWO_{4}$ crystals for the electromagnetic
calorimeter (ECAL). Each crystal measures about $22 \times 22 \, {\rm mm}^{2}$
\cite{stochastic} and covers $0.0175 \times 0.0175 \; (1^{\circ})$ in
the $\Delta \eta -\Delta \phi$ space ($\phi$ being the azimuthal angle). 
For photon
reconstruction, we have used the  ``hybrid'' algorithm
\cite{hybrid}. The first step is to find a seed above a certain
minimum tranverse momentum threshold $P_{T}^{min}$ of 5
GeV\cite{stochastic}. Only electromagnetic objects, i.e., $\gamma,e^{+}$
and $e^{-}$ are chosen as seed. Subsequently, one looks for all
electromagnetic particles around the seed in the $\eta-\phi$ space
where $\Delta\eta$ and $\Delta\phi$ distance from the seed object is
at most 0.09. This extension is equivalent to $10 \times 10$ crystal size in
the CMS detector. The CMS experiment uses $5 \times 5$ crystal size to form an
energy cluster and nearby non-overlapping clusters are merged to
reconstruct a photon candidate. 
However, in our effort to mimic this reconstruction process
at the generator level, we choose to be conservative 
and use only a $10 \times 10$ crystal. 
We define the momentum of a 
photon candidate to be the
vector sum of the momenta of the electromagnetic objects in such a 
crystal. A photon candidate will be either a direct photon or other
electromagnetic obejcts such as $ \pi^{0}\rightarrow
\gamma\gamma,\rho^{0}\rightarrow \gamma\gamma$ etc. Events where the two
highest $E_t$ photons have $\cos(\theta_{\gamma1 \gamma2}) > 0.9$
with $\theta_{\gamma1 \gamma2}$ being the opening angle between the two
photons, are not considered because they could merge into a single
energy cluster in the real detector. We have compared our results with the 
fast detector simulation (FAMOS\cite{famos}) used for CMS experiment and they are
found to be in good agreement. With this algorithm and requiring the
photon to be isolated (to be discussed later), the estimated 
probability of a 
jet faking a photon in $\gamma+jet$ channel is $\sim
10^{-3}- 10^{-4}$. The major sources of fake photons are $\pi^{0}$
$(\sim 81 \%)$, $\eta$ $(\sim 12 \%)$ and $\omega$ $(\sim 3 \%)$, with 
only a small fraction coming from other sources.

\section{Isolation Variables} 
   \label{sec:isolation} 
In a detector, a photon is recognised as a local deposition of
electromagnetic energy in a limited region in the $\eta$--$\phi$ phase
space.  In practice, it is defined as electromagnetic energy contained
in a cone of a given size $R\equiv \sqrt{\Delta\phi^{2} +
\Delta\eta^{2}}$ with no associated tracks. Fake photon signals
arising from a jet can be rejected by requiring either the absence of
charged tracks above a certain minimum transverse momentum ($P_{T
min}^{trk}$)associated with the photon or the absence of additional
energetic particles in an annular cone ($R_{iso}$) around the photon
candidate. We have considered two variables for the isolation purpose
(a) the number of tracks ($N_{trk}$) inside a cone around the photon and
(b) the scalar sum of transverse energy ($E_{TSUM}$) inside a cone around
the photon.

\subsection{Track Isolation}
\begin{figure}[!h]
\vspace*{-2ex}
\scalebox{0.35}{\includegraphics{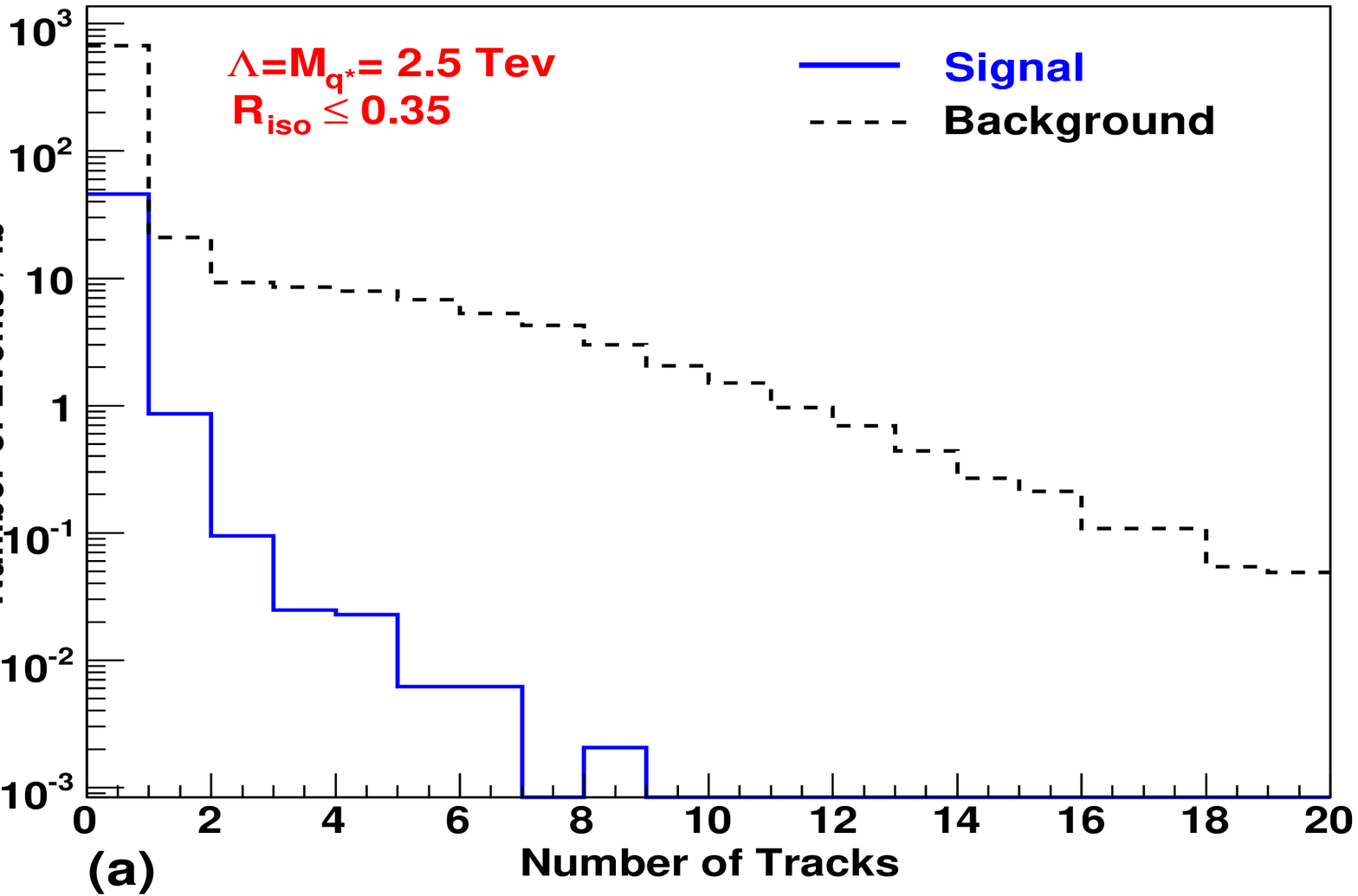}} 
\scalebox{0.35}{\includegraphics{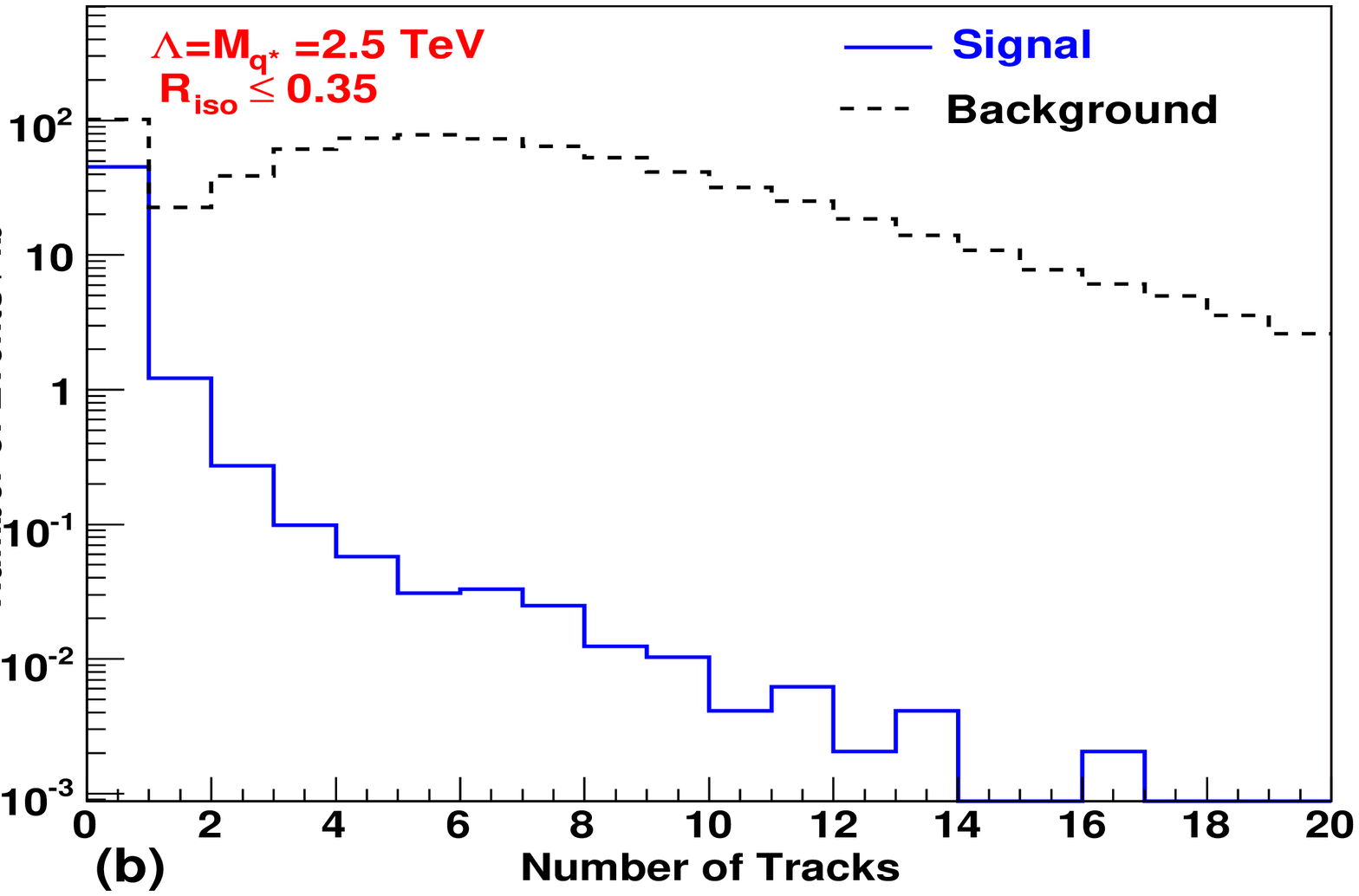}}                             
\caption{Number of tracks for the signal and the background events with $P_{T}^{trk}\geq$ 3.0 GeV pointing (a) leading photon and (b) second leading photon candidates in a cone of size 0.35.}
\label{fig:Trk_N_2lead}
\end{figure}

 We have considered ``stable'' charged particles e.g. $\pi^{\pm}, \,
K^{\pm}, \, e^{\pm}$ and $P^{\pm}$ as tracks.  Of these, $\pi^{\pm}$
alone contribute $\sim 80 \%$ of the total charged tracks. The
contributions from stable charged particles other than the ones
mentioned above are negligible. The distributions of the number of
charged tracks with a requirement on the transverse momentum of the
tracks pointing to either the leading photon or the second leading
photon candidate and within a corresponding cone of size 0.35 are
shown in Fig.~\ref{fig:Trk_N_2lead}. In the signal sample (although we
demonstrate for a particular value of the parameters, the features are
generic), both photon candidates are true photons and hence the
distribution falls very rapidly.  The situation is markedly different
for the background. For a true $\gamma + jet$ event, the second
leading photon is usually the fake one and has a large amount of
hadronic activity around it. Consequently, the distribution (in
Fig.~\ref{fig:Trk_N_2lead}$b$) reaches a maximum around 5--6 tracks
and then falls slowly. To understand the shape of the background
distribution in Fig.~\ref{fig:Trk_N_2lead}$a$, it should be realized
that a small fraction of such events would actually have the fake
photon as the leading one. Since such photons have a large number of
tracks around them, an extended tail as seen in
Fig.~\ref{fig:Trk_N_2lead}$a$ results. The same effect leads to the
rise in the background distribution for the second-leading photon for
$N_{trk} \leq 1$ (Fig.~\ref{fig:Trk_N_2lead}$b$).

  In $pp$ collisions at the LHC, one expects to have a large number of
soft tracks from associated minimum bias and underlying events. The
major sources of tracks in the case of a true photon case are ISR, FSR
and MPI, while the low-$P_{T}^{trk}$ ($< 1.5$ GeV) tracks emanate
mainly from the debris of the colliding protons. If these tracks are
counted, a true isolated photon emitted from a hard $pp$ collision may
also appear non-isolated, thereby reducing the signal efficiency. To
avoid such possibilities, soft tracks are cleaned up by requiring the
tracks to have a $P_{T}$ above a certain minimum threshold ($P_{T
min}^{trk}$). In various CMS studies $P_{T min}^{trk}$ typically
varies between 1-2 GeV \cite{cluster_algo,trkpt,PG}.

\begin{figure}[htbp]
\scalebox{0.45}{\includegraphics{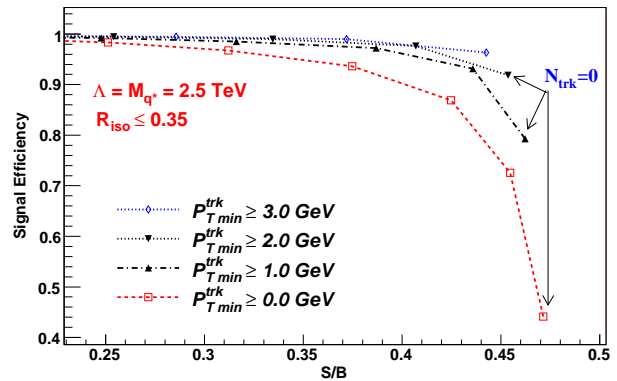}}                                                                      
\caption{ Effect of the minimum threshold for track $P_{T}$ on 
the $S/B$ vs efficiency distribution for the second leading photon.}
\label{fig:trk_min}
\end{figure}

 In this analysis, we have considered several choices for $P_{T
min}^{trk}$, namely 0.0, 1.0, 2.0 and 3.0 GeV respectively, and for
different isolation cone sizes. The signal efficiency and the signal
over background (S/B) ratio were calculated with these choices for
$P_{T min}^{trk}$ and for various $N_{trk}$ possibilities. The
results, for the second leading photon,
 are displayed in Fig.~\ref{fig:trk_min}.  As one can observe, for
$N_{trk}=$ 0, as $P_{T min}^{trk}$ is increased from 1.0 GeV to 3.0
GeV, the signal efficiency increases by more than $15\%$ with only a
small reduction in the S/B ratio. Although, allowing
more tracks in a given cone
size leads to an increase in the signal efficiency, the S/B ratio
decreases drastically (see Fig.~\ref{fig:Trk_N_2lead}).

\begin{figure}[htbp]
\scalebox{0.37}{\includegraphics{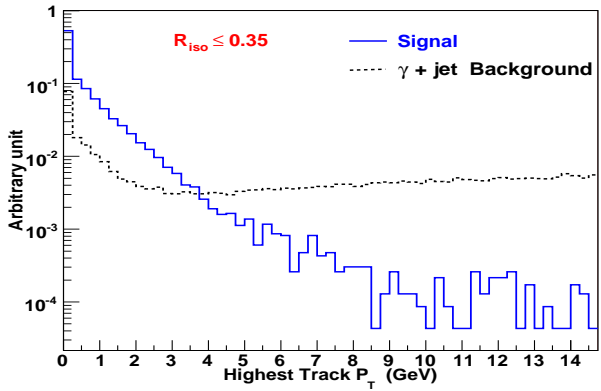}}                                                                               
\caption{Highest track $P_{T}$ around the second leading photon for both 
signal and $\gamma + jet $ background. An isolation cone of size 0.35 
has been used.}
\label{fig:trk_min2}
\end{figure}

Understandably, neither the SM diphoton contribution (whether the Born
or the box-mediated processes) nor the new physics contribution to the
same are affected by the requirement of $N_{trk}$=0.  Only the
$\gamma+jet$ background suffers.  Fig.\ref{fig:trk_min2} shows the
corresponding distribution in $P_{T}$ for the highest transverse
momentum track emanating from the second leading photon. Both the
distributions (signal and background) have been normalized to
unity. Clearly, the background dominates the signal for $P_{T
min}^{trk}>$ 3.5 GeV, thus pointing out a means to reject a large
fraction of the $\gamma+jet $ background. Only those events are
accepted where neither of the photons have an 
associated track with $P_{T}\geq$3.0
GeV within the respective isolation cones (i.e. $N_{trk}$=0 for $P_{T}^{trk}\geq$3.0 GeV). Only the highest $P_{T}$ track is considered
because considering lower $P_{T}$ tracks may affect signal
efficiency. Since this study has been done at the generator level we
have chosen $P_{T min}^{trk}\geq $3.0 GeV.

\subsection{$E_{t}$ Sum Isolation} 

Defined as the cluster of energy inside a cone $\Delta R$ from which
the energy of the photon is subtracted, the variable $E_{T SUM}$ can
be used to discriminate against an event wherein a jet fakes a photon.
Although, in a real detector, $E_{T SUM}$ is separately accounted for
in the electromagnetic and the hadronic calorimeters, due to
limitations of a generator level study, we use a combined $E_{T SUM}$
which is the scalar sum of transverse energy
of the electromagnetic and hadronic particles around the photon
candidate.

\begin{figure}[!h]
\scalebox{0.35}{\includegraphics{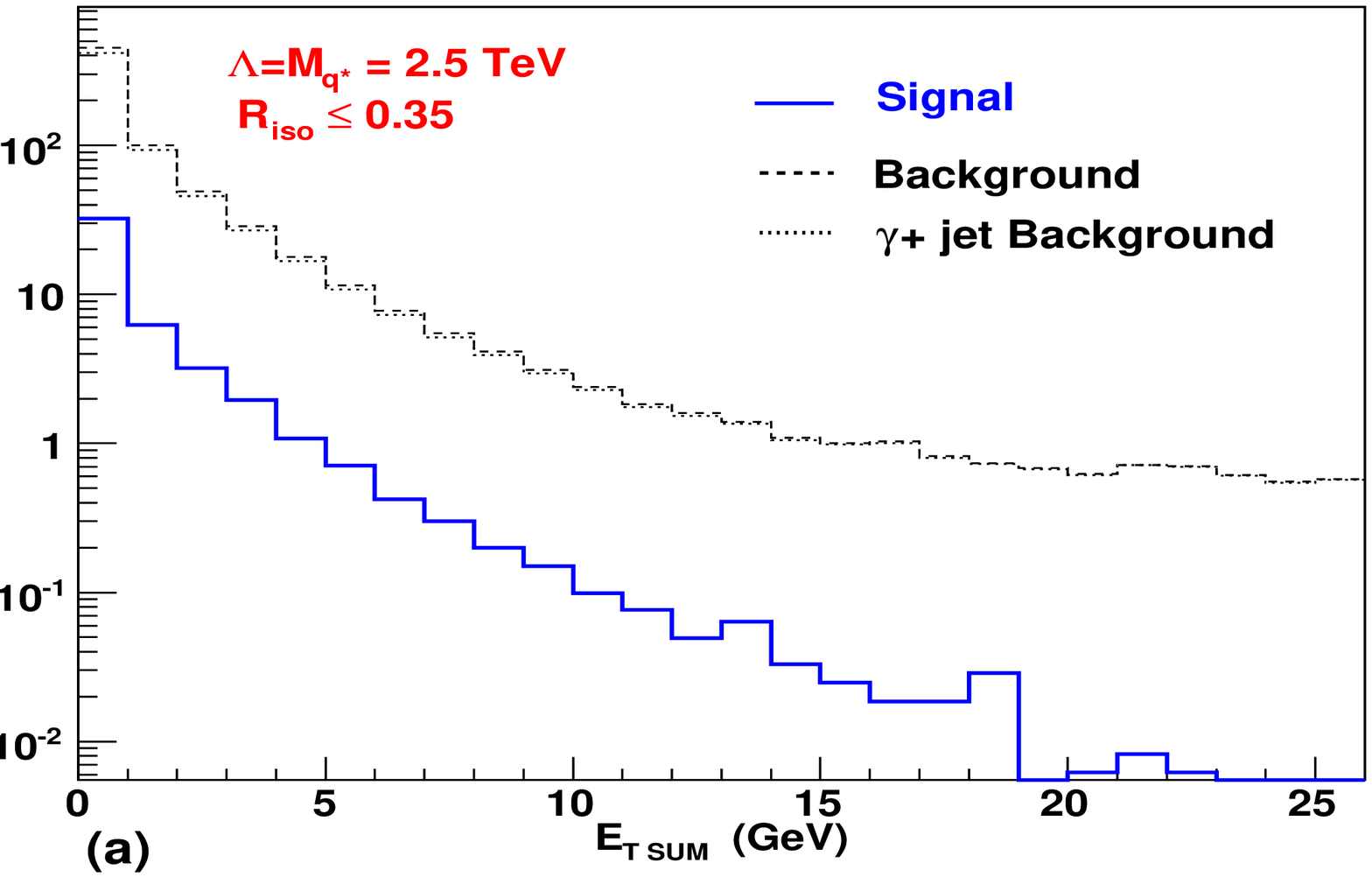}}  
\scalebox{0.35}{\includegraphics{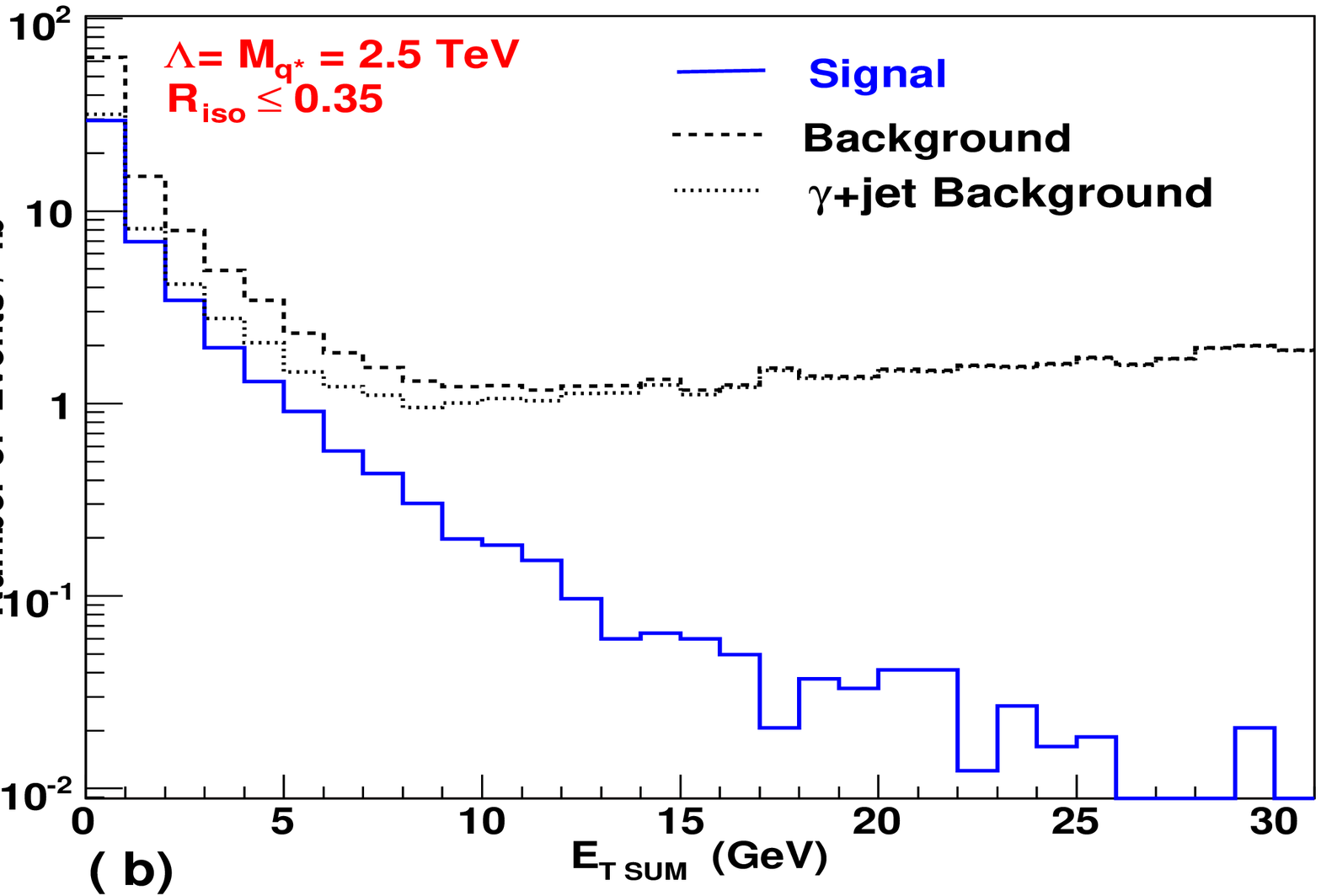}}                             
\caption{$E_{T SUM}$ for the signal and the background events around (a)the  
leading and (b)the next leading photons.}
\label{fig:etsum}
\end{figure}

Fig.~\ref{fig:etsum} shows the normalized $E_{T SUM}$ distributions for the
signal and the backgrounds. The main aim of this study is to
optimize the $E_{T SUM}$ isolation variable so as to reduce the
background from $\gamma+jet$ events. The leading photons, expectedly,
have similar distribution for the signal and the background.  For the
second photon though, the behaviours are very different.  Most of the
$\gamma+jet$ events have $E_{T SUM}>$ 5 GeV and by $E_{T SUM} \gsim$
10 GeV, the S/B ratio is miniscule.

\begin{figure}[!h]
\scalebox{0.45}{\includegraphics{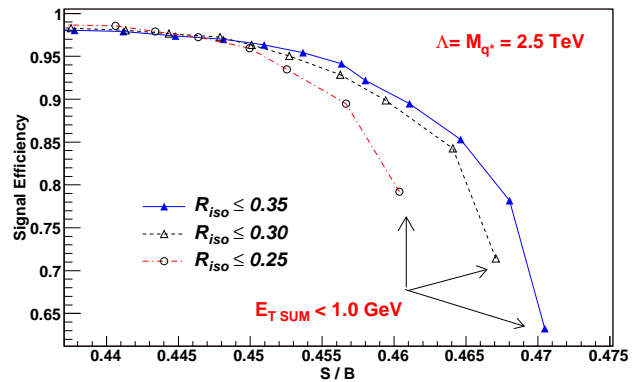}}                             
\caption{Signal efficiency vs. S/B ratio for different cone sizes for 
different choices of the $E_{T SUM}$ threshold around the 
second leading photon.}
\label{fig:etsum2}
\end{figure}

In Fig.~\ref{fig:etsum2}, we show the variation in signal efficiency
and the S/B ratio for different cone sizes around the second leading
photon. Each point corresponds to a different $E_{T SUM }$ threshold, 
varied in steps of 1 GeV beginning with 1.0 GeV. The final choice of the cone
size and the $E_{T SUM}$ threshold depends on the track isolation
efficiency, the signal efficiency, and the S/B ratio.

\subsection{Combined Isolation}

In Table II, we show various combinations of isolation variables for
two different cone sizes. Since we aim to observe an excess of
diphoton production over the SM expectations, it is rather important
to have a large signal efficiency.  We have performed this study for a
large number of $\Lambda-M_{q*}$ points for which the cross section is
slightly larger than $q\bar{q}\rightarrow \gamma\gamma$ production
cross section, or in other words those points for which there will be
only a small excess over the SM background.  Although we have used a
simple approach, it is possible to have other criteria to select
analysis points for the choice of final selection cuts. Based on the
studies detailed above, the final selection cuts are as follows:
\begin{itemize}
\item  $P_{T}^{\gamma1 } \geq 200$ GeV,  $P_{T}^{\gamma2} \geq 200$ GeV;

\item  $|\eta^{\gamma1, \gamma2}| < 2.5 \quad$ \&  
     $\quad|\eta^{\gamma1, \gamma2}| \not \in [1.4442,1.5666]$;

\item  $\cos (\theta_{\gamma1\gamma2}) \leq 0.9$;

\item  $N_{trk}=0$ for $P_{T}^{trk} \geq 3.0$ GeV within $R_{iso}\leq 0.35$; 

\item $E_{T SUM}< 5.0$ GeV  within $R_{iso}\leq 0.35$.
\end{itemize}
After the application of the
fiducial volume and photon $P_{T}$ criteria, the requirement on
angular separation between the photons 
removes only $\sim 1\%$ events.

\begin{table} \caption{Fraction of events
surviving for signal and background after applying isolation cuts on both photons (and the $P_{T}^{\gamma}$ $\&$ $\eta^{\gamma}$ criteria). Also shown is  the $S/B$ ratio.}  
\begin{ruledtabular} \begin{tabular}{ccccccccc} $R_{iso}$ &
$N_{trk}$ & $E_{TSUM}^{max}$ & $P^{trk}_{T min}$ &S\footnotemark[1]
&Born & Box & $\gamma +Jet$ & S\footnotemark[1]$/B$\\
 &  & (GeV) & (GeV) &(\%)  &(\%) &(\%) &(\%) & \\
\hline
0.35&  0&  4.0&  1.5&   75.53&   75.45&  71.86&  0.81&   0.828\\
    &   &     &  2.0&   80.52&   80.40&  76.84&  0.90&   0.824\\
    &   &     &  3.0&   83.33&   83.19&  79.57&  0.96&   0.821\\

    &   &  5.0&  1.5&   77.10&   77.05&  73.60&  0.86&   0.824\\
    &   &     &  2.0&   83.15&   83.05&  79.68&  0.98&   0.818\\
    &   &     &  3.0&   87.18&   87.19&  83.79& 1.09&   0.810\\
\hline
0.30&  0&  4.0&  1.5&   81.20&   80.99&  77.97&  0.97&   0.817\\ 
    &   &     &  2.0&   85.73&   85.59&  82.55&  1.07&   0.811\\
    &   &     &  3.0&   88.49&   88.32&  85.44&  1.15&   0.806\\

    &   &  5.0&  1.5&   82.25&   82.17&  79.10&  1.01&   0.813\\
    &   &     &  2.0&   87.48&   87.45&  85.45&  1.14&   0.805\\
    &   &     &  3.0&   91.30&   91.24&  88.39&  1.26&   0.798\\                

\end{tabular}
\footnotetext[1]{Here $\Lambda = M_{q*} = 2.5$ TeV}
\end{ruledtabular}
\end{table}

\begin{table} \caption{The number of events
surviving for signal and background for $\it L_{int}= 1 fb^{-1}$ after 
applying the final selection criteria.}
\begin{ruledtabular} \begin{tabular}{ccccccccc} $R_{iso}$ &
$N_{trk}$ & $E_{TSUM}^{max}$ & $P^{trk}_{T min}$ & S\footnotemark[1]
& Born & Box & $\gamma +Jet$ & Tot.Back.\\
 &  & (GeV) & (GeV) & & & &  & \\
\hline
0.35&  0&  4.0&  1.5&   46.32&   35.23&  1.90&  5.67&  42.81\\  
    &   &     &  2.0&   49.31&   37.54&  2.03&  6.29&  45.87\\
    &   &     &  3.0&   51.09&   38.85&  2.10&  6.70&  47.66\\

    &   &  5.0&  1.5&   47.33&   35.98& 1.94&  5.99&   43.92\\
    &   &     &  2.0&   50.94&   38.78& 2.11&  6.84&   47.73\\
    &   &     &  3.0&   53.54&   40.71& 2.21&  7.56&   50.49\\
\hline
0.30&  0&  4.0&  1.5&   49.83&   37.82&  2.06&  6.74&  46.62\\ 
    &   &     &  2.0&   52.55&   39.96&  2.18&  7.46&  49.62\\
    &   &     &  3.0&   54.35&   41.24&  2.26&  8.00&  51.51\\
                                                                                    &   &  5.0&  1.5&   50.57&   38.35&  2.09&  7.01&  47.46\\     
    &   &     &  2.0&   53.67&   40.83&  2.23&  7.93&  51.00\\
    &   &     &  3.0&   56.10&   42.60&  2.34&  8.78&  53.75\\

\end{tabular}
\footnotetext[1]{Here $\Lambda = 1.0$ TeV and $M_{q*} = 0.5$ TeV}
\end{ruledtabular}
\end{table}

Table III shows the number of events surviving for signal, Born, box,
$\gamma+jet$ and total background for 1$fb^{-1}$ of integrated
luminosity after applying the final selection criteria.

\begin{figure*}
\scalebox{0.35}{\includegraphics{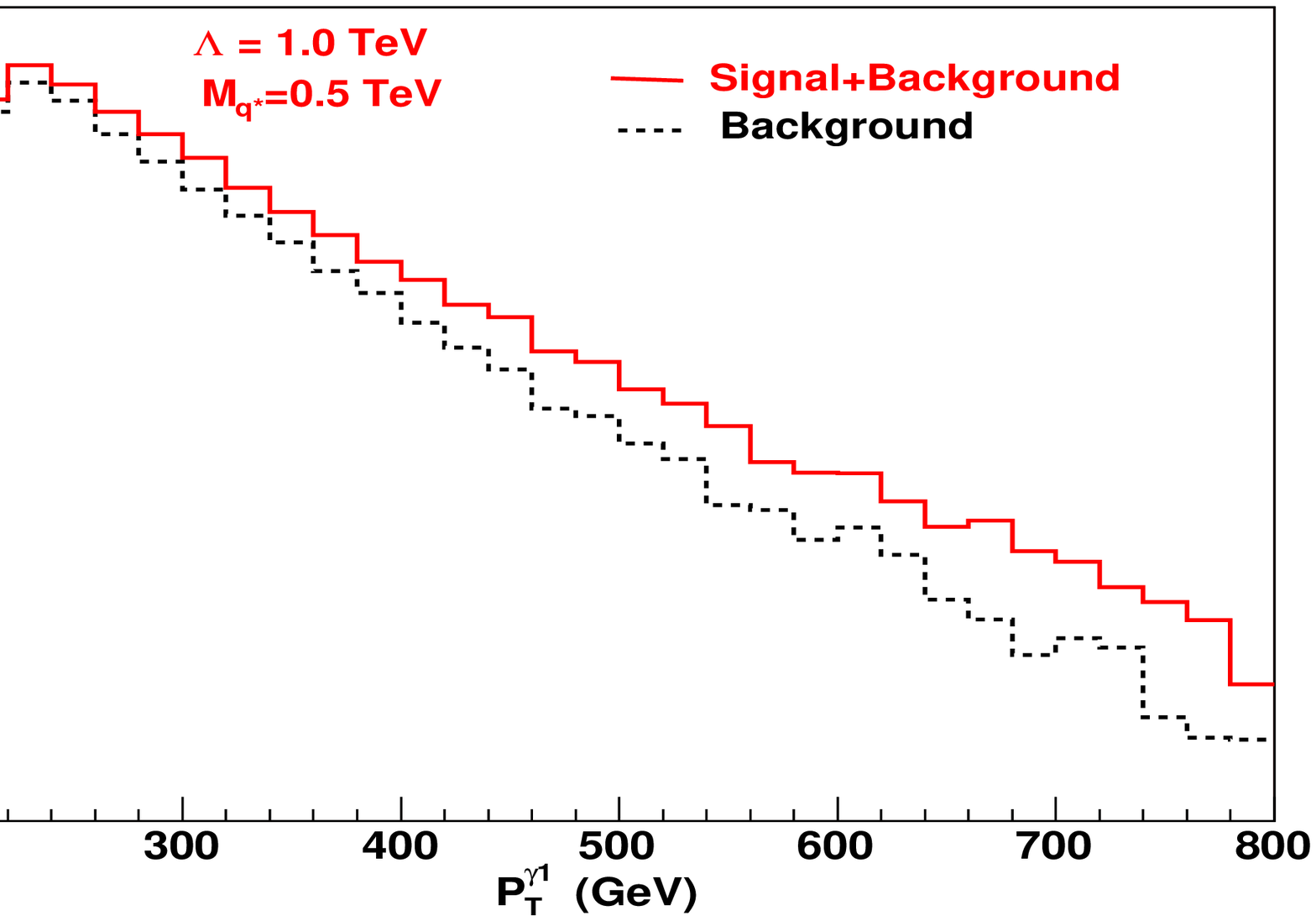}}
\scalebox{0.35}{\includegraphics{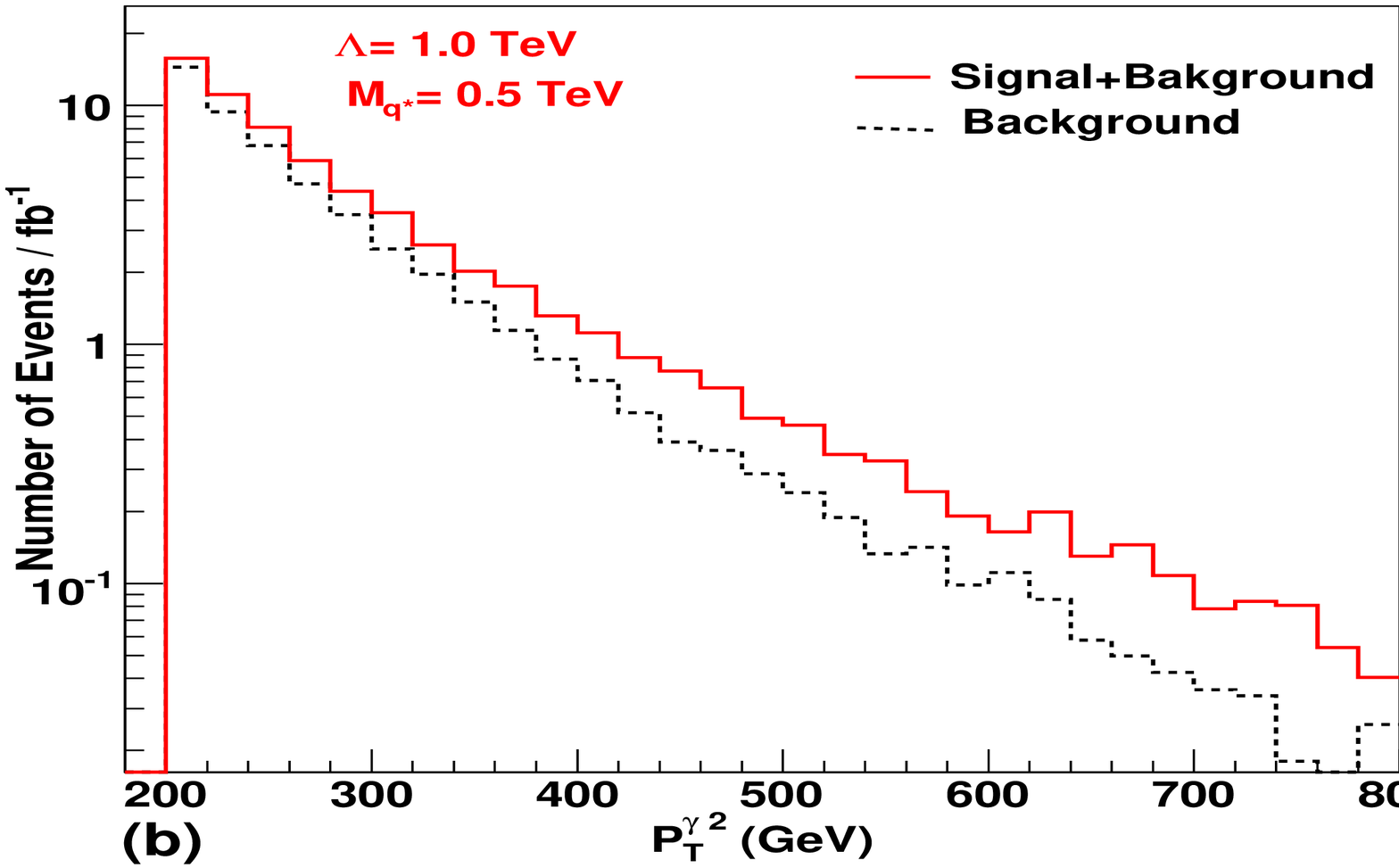}}  
\scalebox{0.35}{\includegraphics{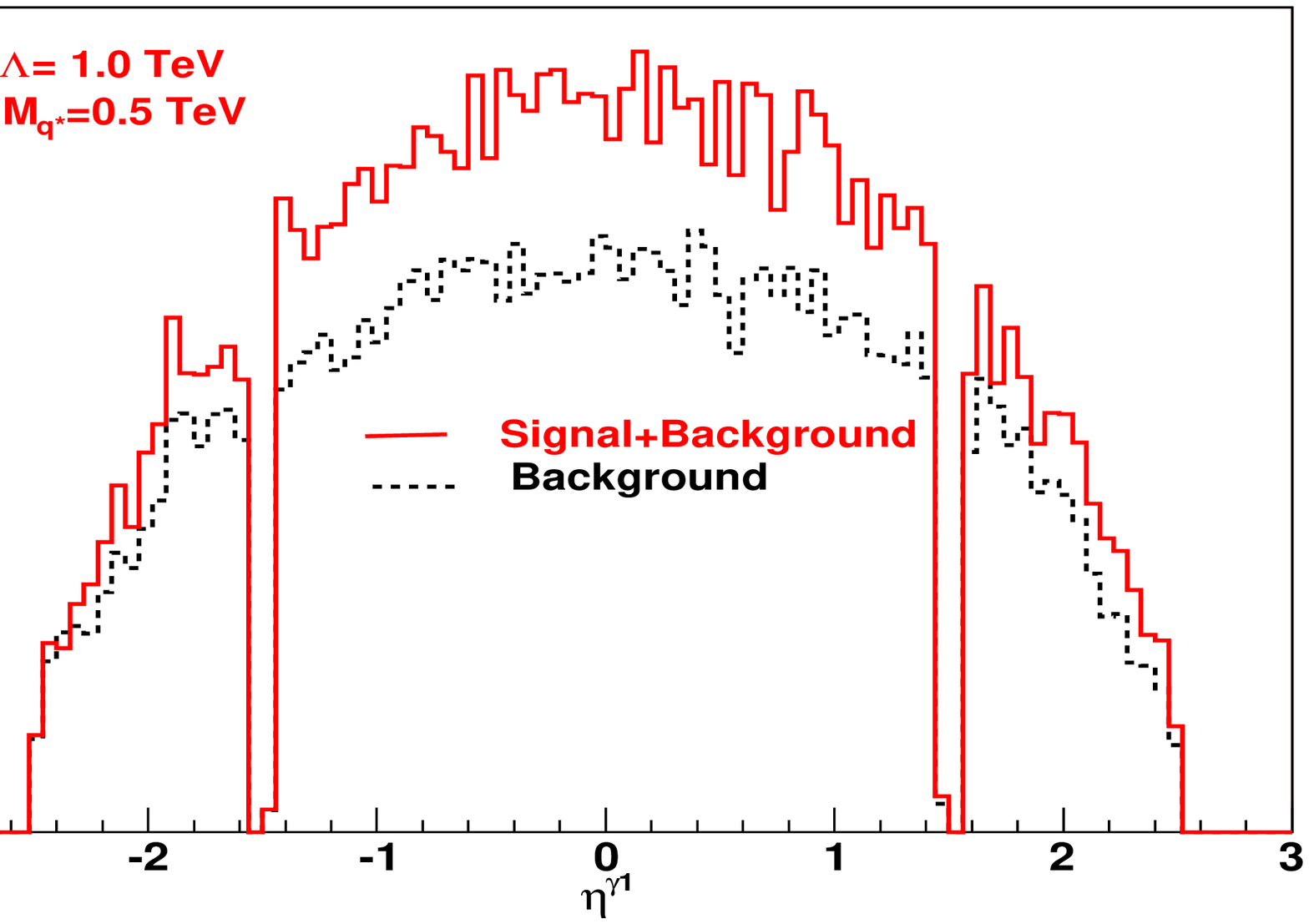}}
\scalebox{0.35}{\includegraphics{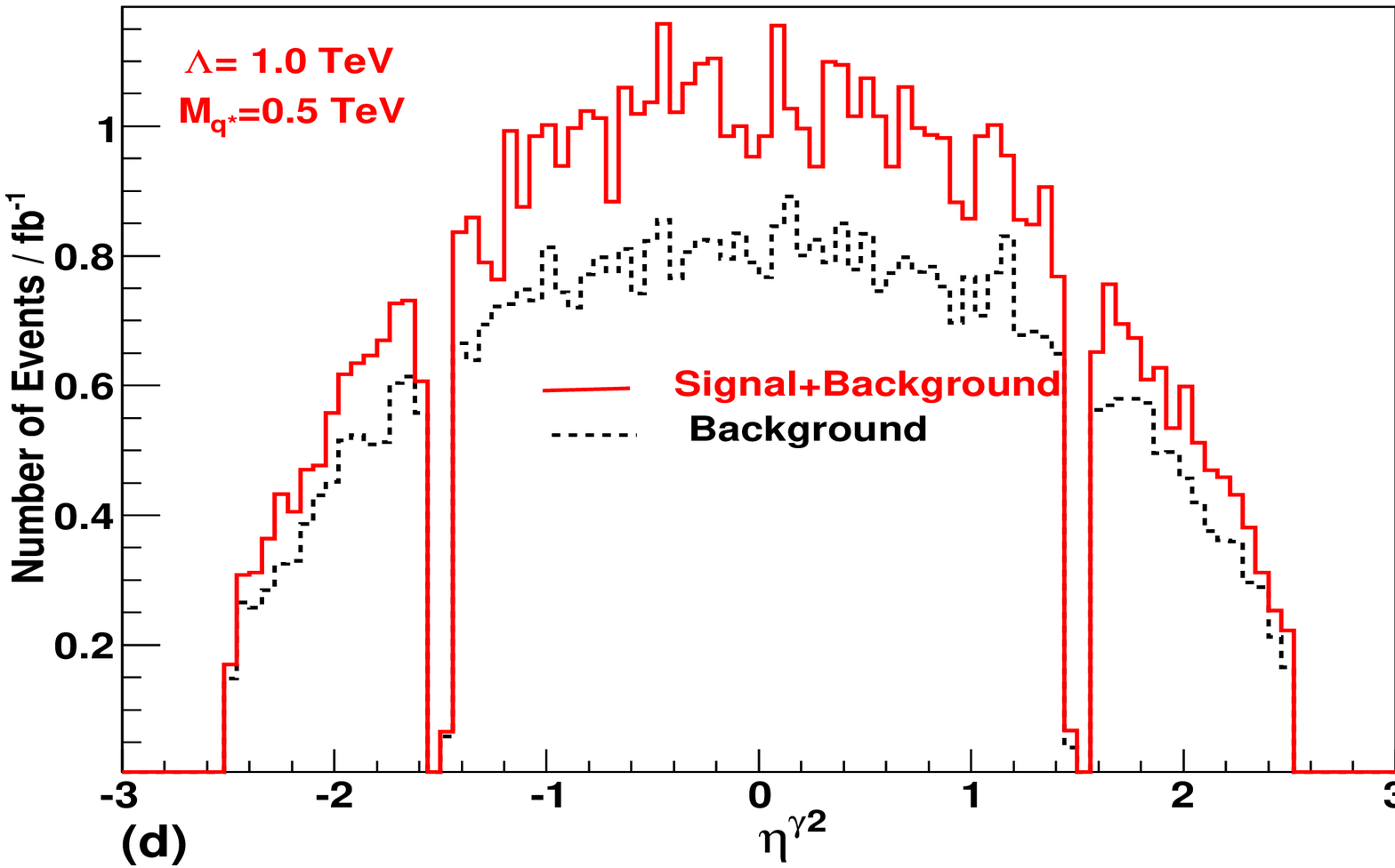}}
\scalebox{0.35}{\includegraphics{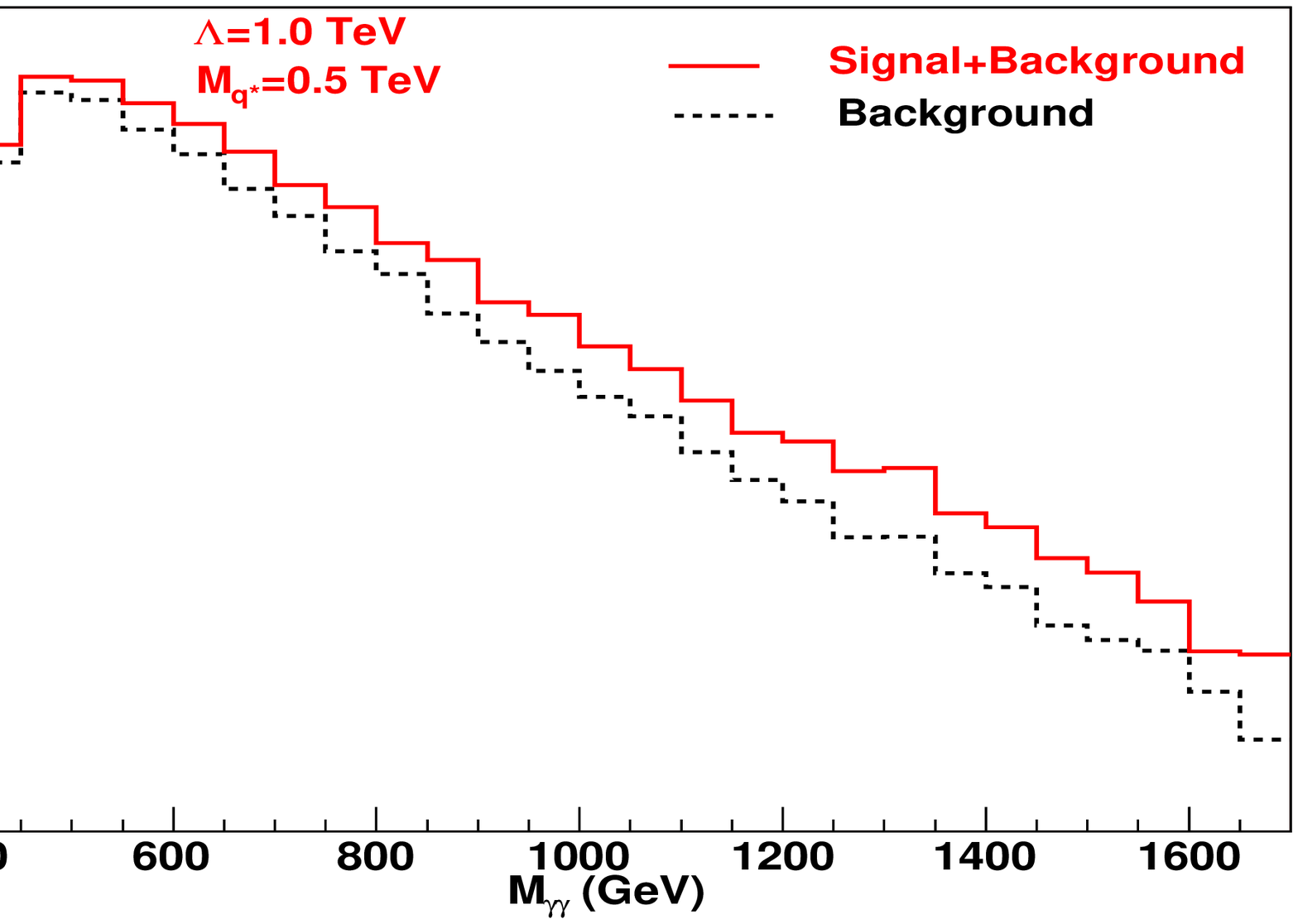}}
\scalebox{0.35}{\includegraphics{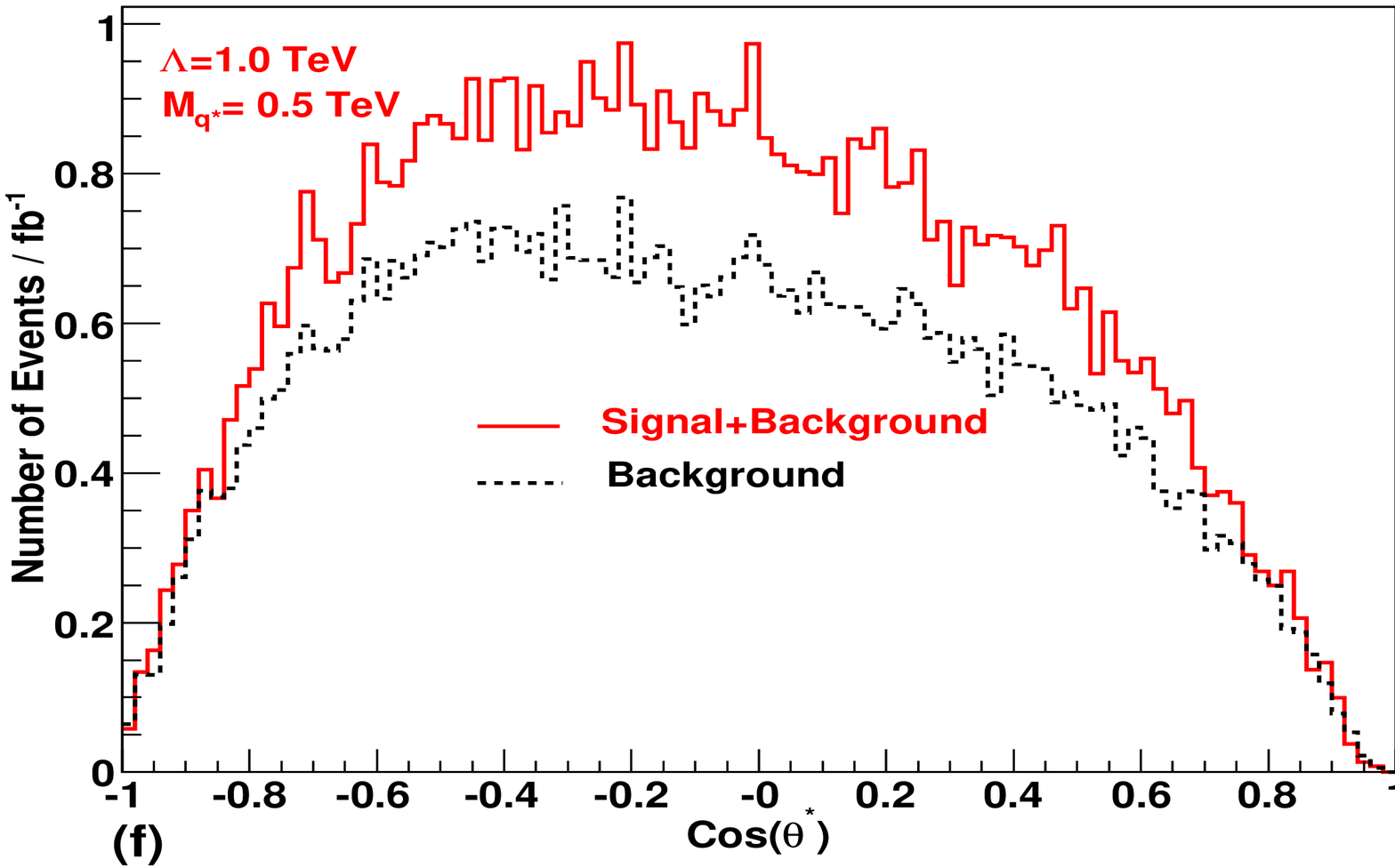}}  
\caption{Kinematic variables after the selection cuts.(a) $P_{T}^{\gamma1}$ distribution,(b) $P_{T}^{\gamma2}$ distribution,(c) $\eta^{\gamma1}$ distribution,(d) $\eta^{\gamma2}$ distribution,(e) $M_{\gamma\gamma}$ distribution and 
(d) $\cos \theta^*$.}
\label{fig:costheta}
\end{figure*}
  Fig. \ref{fig:costheta} shows the distributions for some of the
variables for the generated signal and background events after the
selection requirements are imposed.
In Fig \ref{fig:costheta}f, $\theta^*$ is the
angle between the direction of boost of the diphoton system and each
photon in the diphoton rest frame.

\section{Confidence Level Calculation}

As the q* appears only in the $t$-channel, no resonance peak appears
in the diphoton invariant mass distribution. Rather, a discovery needs to
be made from an observation of enhanced rate in the diphoton channel
as well as differences in the shape of diverse phase space
distributions.  In this analysis, we primarily use the information
contained in the invariant mass distribution to distinguish between
two hypotheses, namely the signal + background hypothesis (S+B) and
the background only (B) hypothesis. We adopt a frequentist approach to
determine the confidence level of rejecting the S+B hypothesis (the
exclusion CL) in the absence of a signal. The histograms shown in
Fig.\ref{fig:costheta}(e) are used to generate two sets of
Gedankenexperiments. To do so, we assume that the content of each bin
in the histograms is Poisson distributed. For every bin, we generate a
Poisson random number, taking the original bin content as the
mean. These Poisson fluctuated random numbers now represent the bin
contents of a histogram which we call the outcome of a single
Gedankenexperiment. One million such Gedankenexperiments are generated
from the S+B histogram (and, similarly, from the B histogram).  From
each of these experiments we calculated our test statistic, namely
\begin{equation}
\chi_{S+B}^{2}= \sum_{i=1}^{n_{bins}}{\frac {(d_{i} -(S+B)_{i})^{2}}{(\sqrt{(S+B)_{i}})^{2}}}
 \label{eq:chi}
\end{equation}
\begin{figure}
\scalebox{0.45}{\includegraphics{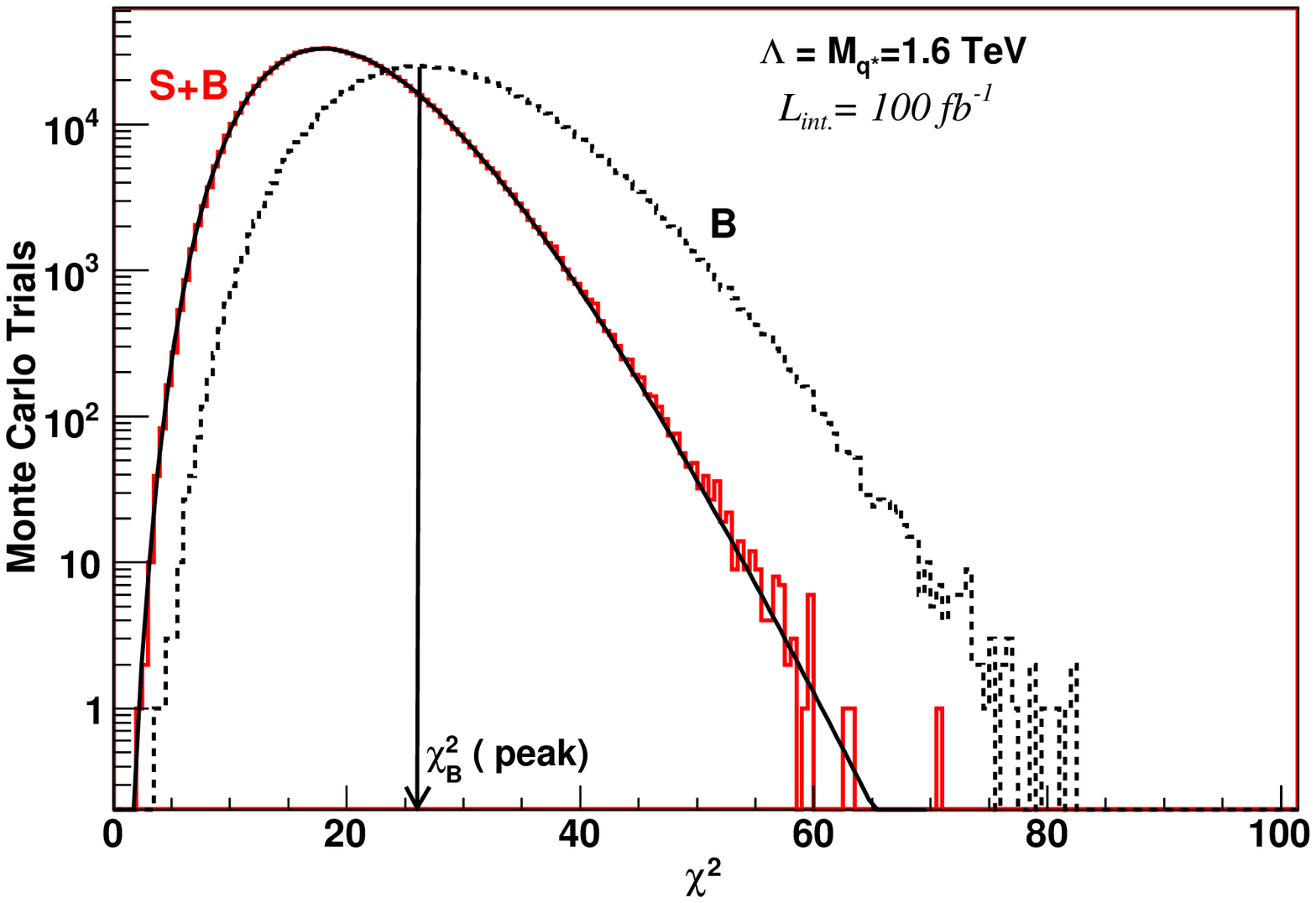}}
\caption{$\chi^{2}$ distribution for $S+B$ and $B$ type hypothesis for a given $\Lambda-M_{q}*$ point with $10^{6}$ MC trials at 100 $fb^{-1}$ of integrated luminosity. Here $S+B$ is fitted with $\chi^{2}$ distribution.}
\label{fig:CHI2}
\end{figure}
(and similarly for $\chi_{B}^{2}$).  Here, $d_{i}$ is the number of
events in the $i^{th}$ bin of the $M_{\gamma\gamma}$ distribution as
generated in a particular Gedankenexperiment and $(S+B)_i$ is the
number of events in the original histogram of $M_{\gamma\gamma}$
obtained from PYTHIA.
 The distribution of $\chi^{2}$ shows how the test statistic will be
distributed over many repeated observations of the mass histogram. In
Fig.~\ref{fig:CHI2}, the solid histogram shows the expected
distribution of $\chi^{2}$ if the S+B hypothesis is true while the
dotted one shows the $\chi^{2}$
distribution if the S+B hypothesis is not true. The most probable value of
$\chi^{2}$ if S+B is false is given by the peak of the $\chi^{2}_{B}$
distribution. The area, $\alpha$ of the $\chi^{2}_{S+B}$ curve to the
right of this value is the probability of seeing a $\chi^{2}$ value
$\geq$ $\chi^{2}_{B}$ (peak) if the S+B hypothesis is true. For every
point in the $(\Lambda,M_{q}*)$ plane satisfying $1-\alpha \geq 99\%$,
the point is rejected at $99\%$ CL.

 In calculating the $\chi^{2}$, only bins with large significance are
used. These have large bin contents and the latter can be safely
assumed to be Gaussian distributed. As a consequence, the $\chi^{2}$
statistic detailed above is equivalent to a log likelihood statistic
for this analysis.

 Since we have used histograms generated from PYTHIA as our input for
the CL extraction there is statistical uncertainty associated with the
procedure, i.e., in a repeat of this MC study the position of the
$\chi^{2}_{B}$ peak will fluctuate, resulting in a different value of
$\alpha$. However at $1- \alpha = 99\%$, this fluctuation is estimated
to be less than $0.5\%$ on either side of the peak.


\section{Results}

\begin{figure}[!h]
\scalebox{0.47}{\includegraphics{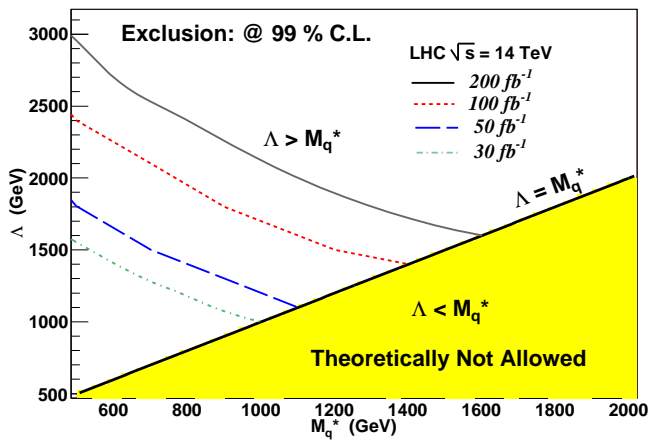}} 
\caption{Achievable exclusion contours in the $\Lambda-M_{q}*$ parameter
space corresponding to different integrated luminosities at the LHC.
The regions below the curves can be ruled out at 99\% C.L.}
\label{fig:result}`
\end{figure}

Fig.~\ref{fig:result} shows the $\Lambda-M_{q}*$ parameter space which
can be excluded for 30, 50, 100 and 200 $fb^{-1}$ of integrated
luminosity. To calculate the limits, we have used the invariant mass
as the discriminating variable. Since the distribution has a long
tail, the analysis has been restricted to $M_{\gamma \gamma} < 1.5$ TeV, so as to
have sufficient events for the considered luminosity. The lower limit
in the $M_{\gamma \gamma}$ was essentially determined by the
requirements on $P_T^\gamma$.

We have checked the stability of the limits and found that the 
99$\%$ CL values suffers only a very small error ($< 0.5 \%$) 
from the uncertainty in the position of the 
$\chi_{B}^{2}$ peak as determined from Monte Carlo trials.
To find the dependence on the choice of kinematical cuts, we
reduced the fiducial volume from $|\eta| <$ 2.5 to $|\eta| <$
1.5. This changes the CL from 98$\%$ to ~99\% CL. Similarly the 
98\% CL limits obtained with $P_{T}^{\gamma} \geq 200$ GeV changes to $99 \% $
CL at $P_{T}^{\gamma} \geq 250$ GeV but at the cost of severe loss in
signal efficiency. Since we have used the deviation of the invariant
mass from the SM prediction as a discriminating variable, we expect to
further improve the limit by combining some other uncorrelated
variables\cite{dvariable}.


\section{Systematics}
As described in the earlier sections, we have 
performed a detailed analysis including a realistic 
simulation of the various detector effects and uncertainties. 
Some systematic uncertainties persist still and, in this section, 
we present an estimation for each of these.
\begin{itemize}
\item
 Choice of PDF: To estimate the uncertainty due to the choice of the
PDF, the cross sections were calculated with different choices of PDFs
and the results obtained compared with those obtained for 
CTEQ6M~\cite{Pumplin:2002vw}. For comparison we used CTEQ5M1, CTEQ5L
and MRST2001. A maximum uncertainty of $\sim$7$\% $ was found when
CTEQ5L was compared to CTEQ6M. For CTEQ5M1 and MRST2001 these values
are $2.3 \%$ and 3.5${\%}$ respectively.

\item
Scale Variation: To estimate this, 
the factorization scale $Q$ (chosen to be $\sqrt{\hat{s}} $ in our analysis)
was varied from in the range $Q^2 \in [\hat s/2, 2\, \hat s]$.
Also used was $Q^2 = P_T^2$. In all these variations, the maximum uncertainty was found to be
 1.6$\%$.

\item Higher-order effects: The SM processes relevant to us have been 
studied in the literature at great length. Most higher order effects 
can be adequately parametrized in the form of a $K$-factor. For 
true diphoton 
production, these are 1.5 (Born process)\cite{k1} and
1.2 (box)~\cite{k2}. For the $\gamma+jet$ events, these are 
1.66 when the quark fragments into a
photon~\cite{k2} and 1.0 when an (almost) isolated $\pi^{0}$ in the 
hadronic jet fakes a photon~\cite{k2}.

For the new physics contribution, the $K$-factor is not known though
(indeed, the very definition could be ambiguous for a
nonrenormalizable theory), and hence we have not used any 
in our analysis. However, in the limit of a very large $M_{q^*}$, 
the new physics effect should be describable in terms of an effective
operator involving quarks and photons and the $K$-factor, in this 
limit, is not expected to be too different from the SM one~\cite{majhi}.

If one 
assumes the signal $K$-factor to be indeed similar to the overall background
one, then the net effect is a scaling of Eq.(\ref{eq:chi}) by a factor of $K$. 
This translates to a modification 
in the separation between the peaks of the two histograms in
Fig.\ref{fig:CHI2} by a factor of $K$ and is
equivalent to an increase in the luminosity by the same factor.
To be conservative, we choose to ignore the consequent improvements in the 
exclusion limits. 

\item Energy resolution: To study the effect of the detector energy
resolution on this analysis, the energy of the photons was smeared with
the stochastic term of the CMS electromagnetic calorimeter energy
resolution\cite{stochastic}. The effect was found to be negligible.

\item Dijet background: Due to limitations in computing resources, we
did not fully simulate the background from jet-jet events. Although the 
dijet cross sections are very large, given the low probability 
of a jet faking a photon (as described earlier in the text), it 
is obviously not very likely that too many such events would 
survive the selection criteria that we have imposed. A parton-level
Monte Carlo calculation readily verified this. 

Even in the corresponding PYTHIA study, it 
 was again observed that the kinematical and isolation cuts reduces this
 background drastically. In a sample of 9000 jet-jet events,  no event
 survives the final selection requirements. However,  with the same survival
 efficiency as for $\gamma+jet $ events (i.e.,$\sim$1 $\%$) and with
 same kinematical and isolation cuts, we expect to have a jet-jet
 background of less than 3.7 events for an integrated luminosity of 1
 $fb^{-1}$. Hence we may safely assume that two photon events from jet-jet
 background will have negligible effect on the final confidence level
 calculation.
\item Luminosity error:
 At the LHC, for an integrated luminosity above 30$fb^{-1}$, the 
error on the measured luminosity is expected to be 3$\%$\cite{sys2}.

We have determined the effect of uncertainty in the theoretical
cross-section on the CL. To get a conservative estimate we lowered the
cross section by $1\%$ and found that 99$\%$ CL changes to 98$\%$
CL.

\end{itemize}

\section{Conclusions}
To summarise, we have investigated the potential of using the
diphoton final state at the LHC in probing possible substructure of
quarks. In any model of quark compositeness, excited states occur
naturally and these couple to the SM counterparts through a
generalised magnetic transition term in an effective
Lagrangian. Consequently, the presence of such states would alter the
diphoton cross section, the extent of which depends on both the mass
$M_{q^*}$ and the compositeness scale $\Lambda$. The deviation
concentrates in the large $p_T$ regime, especially for larger
$M_{q^*}$ and can be substantial. For example, $\Lambda=M_{q}*$=1 TeV
leads to a $\sim$12$\%$ deviation in the cross section (when
restricted to an appropriate part of the phase space as defined in
Section IV).

 Using the photon reconstruction algorithm as used for
the CMS dectector at the LHC, we perform a realistic estimation of the
deviation caused by the excited quark exchange contribution
to the diphoton rate.  We have accounted for
all major backgrounds to evaluate the limits in the $\Lambda-M_{q}*$
parameter space. The possible exclusion limits are very strong and
depend only weakly on the choice of the kinematical cuts.

While direct searches can lead to very strong limits from the 
non-observation of mass peaks, the search strategy outlined here can 
prove to be a complementary tool. In particular, as shown above, this 
mode is sensitive to excited quark masses far above the kinematical 
limit for pair-production (which mainly proceeds through gauge interaction). 
Furthermore, this method is sensitive to the magnetic transition 
coupling ($q^* q \gamma$) in an unambiguous manner free from all other 
couplings and parameters of this essentially complex theory.
\section*{Acknowledgments}
SB and SSC would like to thank Marco Pieri for his comments on
photon algorithm whereas DC would like to thank Samir Ferrag 
for illuminating discussions.
SB and DC acknowledge support from the Department of
Science and Technology(DST), Government of India under project number
SR/S2/RFHEP-05/2006. BCC acknowledge support from the DST,
Government of India under project number SP/S2/K-25/96-V. BCC, SB and SSC gratefully acknowledge the facilities provided by the Center for Detector and Related Software Technology (CDRST), University of Delhi. SSC would like to express gratitude to the Council of Scientific and Industrial Research (CSIR), India for financial assistance and to Prof. R.K. Shivpuri and Prof. Raghuvir Singh for support and encouragement.

\def\Journal#1#2#3#4{{#1} {#2} (#4) #3}
\def\ANNP{\em Ann. Phys. (N.Y.)}
\def\ARNS{\em Ann.~Rev.~Nucl.~Sci.}
\def\EPJC{{\em Eur.~Phys.~J.}{\bf C}}
\def\IJMPE{{\em Int. J. Mod. Phys.} E}
\def\IJMPA{{\em Int. J. Mod. Phys.} A}
\def\JETPLC{{\em Sov. Phys. JETP Lett.} C}
\def\JHEP{{\em J.~High.~E.~Phys.}}
\def\MPLA{{\em Mod.~Phys.~Lett.} A}
\def\NIMA{{\em Nucl.~Instr.~and~Meth.} A}
\def\NIMB{{\em Nucl.~Instr.~and~ Meth.} B}
\def\NCA{{\em Nuovo Cimento} A}
\def\NPA{{\em Nucl. Phys.} A}
\def\NPB{{\em Nucl.~Phys.} B}
\def\NJP{{\em New~J.~Phys.}}
\def\PHYS{{\em Physica}}
\def\PLA{{\em Phys. Lett.} A}
\def\PLB{{\em Phys. Lett.} {\bf B}}
\def\PLD{{\em Phys. Lett.} D}
\def\PL{{\em Phys. Lett.}}
\def\PRL{\em Phys. Rev. Lett.}
\def\PREV{\em Phys. Rev.}
\def\PREP{\em Phys. Rep.}
\def\PRA{{\em Phys. Rev.} A}
\def\PRD{{\em Phys. Rev.} D}
\def\PRC{{\em Phys. Rev.} C}
\def\PRB{{\em Phys. Rev.} B}
\def\PRO{{\em Prog. Theor. Phys.}}
\def\RMP{{\em Rev. Mod. Phys.}}
\def\ZPC{{\em Z. Phys.} C}
\def\ZPA{{\em Z. Phys.} A}

\def\etal{{\em et al.}}
\def\idem{{\em idem}}

\end{document}